\documentclass[aps,pra,reprint,groupedaddress]{revtex4-2}

\usepackage{graphicx}
\usepackage{dcolumn}
\usepackage{bm}

\begin{document}

\title{Comment on "Quantum Fisher information flow and\\ non-Markovian processes of open systems"}

\author{Mihaela Vatasescu}
\email[]{mihaela\_vatasescu@yahoo.com}
\affiliation{Institute of Space Sciences - INFLPR,
MG-23, 77125 Bucharest-Magurele, Romania}
\date{\today}

\begin{abstract}
In [Phys. Rev. A 82, 042103 (2010)], the authors showed that "for a class of the non-Markovian
master equations in time-local forms", the quantum Fisher information (QFI) flow 
can be decomposed into additive subflows
corresponding to different dissipative channels. However, the paper does not specify
the class of non-Markovian time-local master equations for which their analytic
decomposition of the QFI flow is valid. Here we show that several suppositions
have to be made in order to reach the central result of Ref.~\cite{luwsun10}, which appears to be valid for a narrow class of 
density operators $\rho (\theta;t)$ and 
quantum Fisher information $\mathcal{F}(\theta;t)$, and under 
strict conditions on the time-local master equation. More precisely, 
the decomposition of the QFI flow obtained in Ref.~\cite{luwsun10} is valid under two conditions not mentioned in the paper:
(i) $\frac{d}{dt} \left( \frac{\partial \rho}{\partial \theta} \right)=$
$\frac{\partial}{\partial \theta} \left( \frac{d \rho}{dt}  \right)$;
(ii) $\frac{\partial H}{\partial \theta}=0$, $\frac{\partial \gamma_i}{\partial \theta}=0$, 
$\frac{\partial A_i}{\partial \theta}=0$, meaning that the Hamiltonian $H(t)$, the decay rates $\gamma_i(t)$, and the Lindblad operators $A_i(t)$ appearing in the non-Markovian time-local master equation have to not depend on the parameter $\theta$ about which the quantum Fisher information is defined.

\end{abstract}


\maketitle

In Ref.~\cite{luwsun10}, the authors established a remarkable relation between the
quantum Fisher information (QFI) flow and the non-Markovian dynamics of an open system obeying a
time-local master equation. It is stated that "for a class of the non-Markovian
master equations in time-local forms", the QFI flow can be decomposed into additive subflows
corresponding to different dissipative channels characterized by the decay rates
$\gamma_i(t)$. However, the paper does not specify the characteristics of the non-Markovian
time-local master equations allowing to obtain their analytic result. In this Comment we show that this decomposition of the QFI flow (the central proposition of Ref.~\cite{luwsun10}) is valid under 
certain conditions which are not formulated in the paper.  

{\it I. Conditions allowing to obtain the analytic result contained in the central proposition of \cite{luwsun10}. Density operator $\rho(\theta;t)$ and quantum Fisher information $\mathcal{F}(\theta;t)$ with only explicit time dependence.}
Ref.~\cite{luwsun10} uses the partial derivative $\partial / \partial t$ with respect to time to write the 
non-Markovian time-local master equation, as well as to define the QFI flow. This choice is not explained in the paper, but in our view it indicates that the authors selected a narrow class of density operators and QFI having only explicit time dependence.  For this class, we will show below that the main result
of Ref.~\cite{luwsun10} is obtained under two conditions not mentioned in the paper. 

Ref.~\cite{luwsun10} considers an open quantum system whose dynamics is described 
by the time-local master equation 
\begin{eqnarray}
 \frac{\partial \rho(t)}{\partial t}= {\cal K}(t) \rho(t),
\label{drhodt1}
\end{eqnarray}
where $\rho(t)$ is the density operator of the open quantum system, and
 ${\cal K}(t)$ is the generator
of the non-Markovian time-local master equation in the extended Gorini-Kossakowski-Sudarshan-Lindblad (GKSL) form \cite{bookBP2002}:
\begin{eqnarray}
{\cal K}(t) \rho = -i [ H, \rho]
+ \sum_{i} \gamma_i \left[ A_i \rho A^+_i - \frac{1}{2} \{ A^+_iA_i, \rho \} \right],
\label{drhodt2}
\end{eqnarray}
where the Hermitian Hamiltonian $H(t)$, decay rates $\gamma_i(t)$, and Lindblad operators
$A_i(t)$ are time-dependent. Moreover, the decay rates $\gamma_i(t)$ may be negative, unveiling the non-Markovian
character of the dynamics \cite{rivplenio2014,*breuer16}. 

The paper \cite{luwsun10} aims to establish a connection between the  QFI flow and the
non-Markovian character of the quantum dynamics  manifested through the appearance of 
negative decay rates $\gamma_i(t)$ in  Eqs.~(\ref{drhodt1},\ref{drhodt2}). 

Taking $\theta$ as the parameter to be estimated from the density operator $\rho(\theta;t)$,
the corresponding QFI is defined as \cite{helstrom,holevo}
\begin{eqnarray}
\mathcal{F}(\theta;t) := \text{Tr}[ L^2(\theta;t)\rho(\theta;t)],
\label{qFi}
\end{eqnarray}
where $L(\theta;t)$ is the Symmetric Logarithmic Derivative (SLD) defined as the Hermitian
operator satisfying the equation
\begin{equation}
 \frac{\partial \rho(\theta;t)}{\partial \theta }=  \frac{\rho(\theta;t) L(\theta;t) + L(\theta;t) \rho(\theta;t)}{2}.
\label{SLDl}
\end{equation}

Ref.~\cite{luwsun10} uses the above equations to obtain an expression of the QFI flow,
defined as the change rate $\mathcal{I} := \partial \mathcal{F} / \partial t$, claiming that
for an open quantum system whose dynamics is described by  Eq.~(\ref{drhodt1}), the QFI flow
can be explicitly written as a sum of subflows 
$\mathcal{I} = \sum_i \mathcal{I}_i = \sum_i \gamma_i \mathcal{J}_i$:
\begin{equation}
\frac{\partial \mathcal{F}}{\partial t} = 
- \sum_i \gamma_i \text{Tr} \{ \rho [L,A_i]^+[L,A_i] \},
\label{dFdtLu10}
\end{equation}
with $\mathcal{J}_i := -\text{Tr} \{ \rho [L,A_i]^+[L,A_i] \} \le 0$. Consequently, the sign of the subflow $\mathcal{I}_i$ is determined by the sign of the decay rate $\gamma_i$, such that to a negative  $\gamma_i(t)$ is associated a positive subflow $\mathcal{I}_i(t)$. This specific relation between the QFI flow and the decoherence rates installs the QFI flow as a remarkable marker of non-Markovian behavior.

We will show that Eq.~(\ref{dFdtLu10}) is not generally valid for the time-local equation (\ref{drhodt1}),
 being obtained in specific conditions not mentioned in the paper \cite{luwsun10}.
To this aim, we will detail the proof briefly sketched in the paper, showing the steps where mathematical suppositions have to be made in order to obtain Eq.~(\ref{dFdtLu10}).

Before proceeding to the proof, we have to observe that it seems reasonable to consider
that the Hamiltonian $H$, the decay rates $\gamma_i$, and the Lindblad operators $A_i$ implied in  Eq.~(\ref{drhodt2}) depend not only on time, but also on the inference parameter $\theta$:  $H(\theta;t)$, $\gamma_i(\theta;t)$, $A_i(\theta;t)$. 
From the definition (\ref{qFi}):
\begin{eqnarray}
\frac{\partial \mathcal{F}}{\partial t} = \text{Tr} \left( \frac{\partial L^2}{ \partial t} \rho  \right)   
+  \text{Tr} \left( L^2  \frac{\partial  \rho }{ \partial t} \right).   
\label{dFdt0}
\end{eqnarray}
If the term  $\text{Tr} \left( \frac{\partial L^2}{ \partial t} \rho  \right)$ is developed using the partial derivative of Eq.~(\ref{SLDl}) with respect to time \cite{luwsun10}, the QFI flow becomes
\begin{eqnarray}
\frac{\partial \mathcal{F}}{\partial t} = 
\text{Tr} \left\{ L \left[ 2 \frac{\partial}{\partial t} \left( \frac{\partial \rho }{ \partial \theta} \right) 
- L \frac{\partial \rho }{ \partial t}  \right]   \right\}.
\label{dFdt1}
\end{eqnarray}

At this stage, our observation is that, in order to define the operator $\cal{L}$$:= L(2 \partial / \partial \theta -L)$ \cite{luwsun10}, which allows to write the  QFI flow as \cite{luwsun10}
\begin{eqnarray}
\frac{\partial \mathcal{F}}{\partial t} = \text{Tr} \left[\mathcal{L} \left(  \frac{\partial \rho }{ \partial t}\right) \right] \label{cond1a} \\
=  \text{Tr} \left[\mathcal{L}  {\cal K}(t) \rho(t) \right],
\label{cond1b}
\end{eqnarray}
condition (i):
\begin{equation}
\frac{\partial}{\partial t} \left( \frac{\partial \rho }{ \partial \theta} \right) =
\frac{\partial}{\partial \theta} \left( \frac{\partial \rho }{ \partial t} \right)
\label{cond1}
\end{equation}
 has to be fulfilled. 
This relation is known to be valid if the second partial derivatives 
$\frac{\partial^2 \rho}{\partial t \partial \theta  }$, $\frac{\partial^2 \rho}{\partial \theta \partial t }$ are continuous \cite{TaylorMann}.
Eq.~(\ref{cond1}), named here condition (i), provides the "linearity of the QFI flow equation (\ref{cond1a}) with respect to $\frac{\partial \rho }{ \partial t}$" \cite{luwsun10},
on which the main result of the paper \cite{luwsun10} is grounded.

Using Eqs.~(\ref{cond1b},\ref{drhodt2}) one obtains
\begin{eqnarray}
 \frac{\partial \mathcal{F}}{\partial t} = -i \text{Tr} \{ \mathcal{L} [ H, \rho] \} 
+ \sum_{i} \text{Tr} \left[ \mathcal{L} \left( \gamma_i  A_i \rho A^+_i \right)   \right] \nonumber\\
- \frac{1}{2} \sum_{i} \text{Tr} \left[ \mathcal{L} \left( \gamma_i \{ A^+_iA_i, \rho \} \right) \right].
\label{cond1c}
\end{eqnarray}

After the necessary algebra, one finds the following expressions for the three significant terms on the right side of Eq.~(\ref{cond1c}):
\begin{equation}
\text{Tr} \{ \mathcal{L} [ H, \rho] \} = 2 \text{Tr} 
\left( L \left[ \frac{\partial H}{\partial \theta}, \rho \right]  \right),
\label{termH}
\end{equation}
\begin{eqnarray}
\text{Tr} \left[ \mathcal{L} \left( \gamma_i  A_i \rho A^+_i \right)   \right] =
\text{Tr} T^i_1 \left( \frac{\partial \gamma_i}{\partial \theta}, \frac{\partial A_i}{\partial \theta},
  \frac{\partial A^+_i}{\partial \theta}, \gamma_i, L, A_i,A^+_i, \rho \right) \nonumber\\
+ \gamma_i \text{Tr} ( L A_i L \rho A_i^+ + 
L A_i \rho L A_i^+ -L^2 A_i \rho A_i^+),\nonumber\\
\label{term2}
\end{eqnarray}
where $T^i_1 \left( \frac{\partial \gamma_i}{\partial \theta}, \frac{\partial A_i}{\partial \theta},
  \frac{\partial A^+_i}{\partial \theta}, \gamma_i, ... \right)$ designates the sum of all terms containing 
$\frac{\partial \gamma_i}{\partial \theta}$, $\frac{\partial A_i}{\partial \theta}$,
  $\frac{\partial A^+_i}{\partial \theta}$, $\gamma_i$, and depending also on other operators, as it appears in Eq.~({\ref{term2}}).
\begin{eqnarray}
\text{Tr} \left[ \mathcal{L} \left( \gamma_i \{ A^+_i A_i, \rho \} \right) \right] =
\text{Tr} T^i_2 \left( \frac{\partial \gamma_i}{\partial \theta},   \frac{\partial (A^+_iA_i)}{\partial \theta},
  A^+_i A_i, \rho, L  \right) \nonumber\\
+
 \gamma_i \text{Tr} \left( 2 L \left\{ A^+_i A_i,  \frac{\partial \rho}{\partial \theta}  \right\}
- L^2  A^+_i A_i \rho - L^2 \rho A^+_i A_i \right),\nonumber\\
\label{term3}
\end{eqnarray}
with $T^i_2 \left( \frac{\partial \gamma_i}{\partial \theta},  \frac{\partial (A^+_iA_i)}{\partial \theta}, ... \right)$  the sum of the terms containing $\frac{\partial \gamma_i}{\partial \theta}$,
$\frac{\partial (A^+_i A_i)}{\partial \theta}$, and depending on the operators specified in parenthesis in Eq.~({\ref{term3}}).

Finally, Eq.~(\ref{cond1c}) becomes
\begin{eqnarray}
\frac{\partial \mathcal{F}}{\partial t} = 
-2i \text{Tr} \left( L \left[ \frac{\partial H}{\partial \theta}, \rho \right]  \right)
+ \sum_{i} T^i_1 \left( \frac{\partial \gamma_i}{\partial \theta}, \frac{\partial A_i}{\partial \theta},
  \frac{\partial A^+_i}{\partial \theta}, \gamma_i ... \right) \nonumber\\
+ \sum_{i} T^i_2 \left( \frac{\partial \gamma_i}{\partial \theta},  \frac{\partial (A^+_iA_i)}{\partial \theta},
  ... \right)
- \sum_i \gamma_i \text{Tr} \{ \rho [L,A_i]^+[L,A_i] \}. \nonumber\\
\label{dFdt}
\end{eqnarray}

Eq.~(\ref{dFdt}) is the complete expression of the QFI flow obtained from Eq.~(\ref{cond1b}). Only if 
a second condition (ii) is fulfilled, namely
\begin{eqnarray}
\frac{\partial H}{\partial \theta}=0, \frac{\partial \gamma_i}{\partial \theta}=0, 
\frac{\partial A_i}{\partial \theta}=0, 
\label{cond2}
\end{eqnarray}
it becomes possible to write Eq.~(\ref{dFdtLu10}), which is the central proposition of Ref.~\cite{luwsun10}.

{\it II. Density operators and quantum Fisher information with implicit time dependence.}
Let us observe that, in general, both the density operator $\rho(\theta, \lambda_i(t),t)$  and the QFI $\mathcal{F}$$_{\theta}$$(\theta,\lambda_i(t),t)$ may depend implicitly on time through other functions $\lambda_i(t)$ ($\theta$ being considered at a fixed value in local quantum estimation theory \cite{paris09}), or may not have an explicit time dependence. 
In the general case, the total time derivative  $d / dt$ has to be used to write the form
 $d \rho(t) /d t = {\cal K}(t) \rho(t)$
 of the time-local master equation \cite{bookBP2002}, and to define the QFI flow, 
$\mathcal{I} := d \mathcal{F}_{\theta} / d t$. In this case,
the linearity of the QFI flow  with respect to $d \rho / d t$ can be obtained
with a more demanding condition (i):
\begin{equation}
\frac{d}{dt} \left( \frac{\partial \rho}{\partial \theta} \right)=
  \frac{\partial}{\partial \theta} \left( \frac{d \rho}{dt}  \right),
\label{cond1dt}
\end{equation}
implying not only the symmetry of the second partial derivatives, but conditions on the implicit time dependence. Obviously,
condition (ii) rests mandatory to obtain the simplicity of the decomposition (\ref{dFdtLu10}) for
 $d \mathcal{F}_{\theta} / d t$.

In summary, the comment unveils the specific conditions allowing to obtain the decomposition of the QFI flow given in the central proposition of Ref.~\cite{luwsun10}. 
 We have discussed the  specificity of the framework (the use of the partial derivative with respect to time to write the non-Markovian time-local master equation, as well as to define the QFI flow)
used in the paper \cite{luwsun10},
implying  density operators $\rho(\theta,t)$ and QFI $\mathcal{F}$$_{\theta}$$(\theta,t)$ characterized by only explicit time dependence.
For this class, we have shown that the analytic result (Eq.~(\ref{dFdtLu10}))
of Ref.~\cite{luwsun10} is obtained under two conditions not mentioned in the paper: 
(i) $\frac{\partial}{\partial t} \left( \frac{\partial \rho }{ \partial \theta} \right) =
\frac{\partial}{\partial \theta} \left( \frac{\partial \rho }{ \partial t} \right)$, valid if the second partial derivatives are continuous;
(ii) The Hamiltonian $H(t)$, decay rates $\gamma_i(t)$, and Lindblad operators $A_i(t)$ appearing in the time local master equation do not depend on the  parameter $\theta$ about which the QFI $\mathcal{F}$$_{\theta}$ is defined, which means that
$\frac{\partial H}{\partial \theta}=0$, $\frac{\partial \gamma_i}{\partial \theta}=0$, 
$\frac{\partial A_i}{\partial \theta}=0$. 
Looking at the extension of this result to density operators and QFI
which may depend implicitly on time,
we observe that the decomposition (\ref{dFdtLu10}) can be maintained for a QFI flow 
 $d \mathcal{F}_{\theta} / d t$ defined using the total time derivative  $d/dt$, 
provided that a more demanding first condition
$\frac{d}{dt} \left( \frac{\partial \rho}{\partial \theta} \right)=$
  $\frac{\partial}{\partial \theta} \left( \frac{d \rho}{dt}  \right)$ holds, while condition (ii)
rests unchanged.

\begin{acknowledgments}
Financial  support  from  the  LAPLAS  6  program  of  the
Romanian National Authority for Scientific Research 
(CNCS-UEFISCDI) is gratefully acknowledged.
\end{acknowledgments}

\bibliography{commentMV}

\end{document}